\documentclass[11pt]{article}
\newcommand{\be}{\begin{equation}}
\newcommand{\ee}{\end{equation}}
\newcommand{\BE}{\begin{eqnarray}}
\newcommand{\EE}{\end{eqnarray}}
\newcommand{\nn}{\nonumber}
\newcommand{\AAV}{Aharonov, Anandan and Vaidman}
\newcommand{\midd}[1]{\langle #1 \rangle}
\newcommand{\forget}[1]{}

\newcommand{\ket}[1]{| #1 \rangle}
\newcommand{\bra}[1]{\langle #1 |}
\newcommand{\proj}[1]{| #1 \rangle \langle #1 |}
\newcommand{\inpr}[2]{\langle #1 | #2 \rangle}
\def\openone{\leavevmode\hbox{\normalsize1\kern-3.8pt\large1}}
\usepackage{a4,epsfig}
    
\title{
How to protect the interpretation of the wave function 
against protective  measurements}
\author{Jos Uffink
\\
Institute for
 History and Foundations of Mathematics and Science,\\
 University of Utrecht, P.O.Box 80.000, 3508 TA Utrecht, The
 Netherlands\\ ({\tt uffink@phys.uu.nl})\\
}
\begin{document}
\maketitle

\begin{abstract}
A new type of procedures, called protective measurements, has
been proposed by \AAV. These authors argue that a protective measurement 
allows the determination of arbitrary
observables of a single quantum system and claim that this favors a
realistic interpretation of the quantum state. 
This paper proves that only observables that commute with the system's
Hamiltonian can be measured protectively. It is argued that this
restriction saves the coherence of alternative interpretations.
\\[1\baselineskip]
 \end{abstract}
\section{Introduction}

Recent work of Aharonov, Anandan and Vaidman \cite{AV,AAV,AAV2} introduced
 a new type of procedures in quantum mechanics, which they
called `protective measurements'.
 In these procedures, one can measure, under certain conditions and for a
specific set of states, the expectation value of an arbitrary observable
of an individual system. Remarkably, such expectation values are
obtained
while avoiding the subsequent entanglement of the states of the system and
the apparatus, even if the system was initially not in an eigenstate of
the measured observable.
	 In this respect, the protective measurement is very different
from the more well-known von Neumann measurement procedure.

\AAV{} attribute this feature of protective measurements to a physical
manifestation of the wave function of the system \cite[p.4619]{AAV}. They
claim that by means of these measurements one can directly \emph{observe}
the wave function (or quantum state) of an individual system, and conclude
from this that this quantum state should be given an ontological
interpretation:  if it is possible to observe the state of an individual
system, it must correspond to a real property of this system.

This conclusion stands in sharp contrast to  received opinion.
 To be sure, there is no consensus in the literature on the interpretation
of the quantum state. But there does seem to be consensus that the state
of an individual system is unobservable (i.e.\ not empirically
accessible).  In fact, it is only because of this generally shared opinion
that so many different views on its meaning can peacefully coexist
today.
   However, according to the above claims, we can decide the issue of the
interpretation of the wave function by exploiting theoretical
possibilities for measurement allowed by quantum theory itself. This
poses a serious threat for many of these interpretations.

Several critical discussions of these exciting claims have already been
published (See \cite{Schwinger}--\cite{Dass}).
 In this paper I will address an issue which, to my knowledge, has not
previously been dealt with.
 I show that the conditions assumed  in a protective
measurement imply that the observable that is being
measured  commutes with the Hamiltonian of the system.
 While this limitation still allows the possibility of a unique
determination of the quantum state within the specific set considered in a
protective measurement, it undermines the claim that one sees its direct
physical manifestation. I conclude that the threatened interpretations of
the wave function can be saved in the face of protective measurements.

This paper is organized as follows. In section 2 the theoretical
background of protective measurements is reviewed. Section 3 discusses its
consequences for the interpretation of quantum mechanics in more detail.
In section 4, I prove the limitations on the observables that can be
measured protectively. Section 5 provides an explanation of the mechanism
of protective measurement that does not involve the manifestation of the
wave function and applies this to a thought experiment of
\AAV. Section 6 argues that alternative interpretations are not endangered
by protective measurements, and discusses some possible objections to this
conclusion.
 
\section{Protective measurements}
The notion of a protective measurement is introduced 
by means of a concrete measurement model.
Consider a system  $S$ in interaction with some measurement apparatus
$A$ and let ${ \cal H}_S \otimes{ \cal H}_A$ be their composite 
Hilbert space.
Assume that the total Hamiltonian of this composite system is of the form
  \be H_{\rm tot}(t)  = H_S +H_A +  g(t) H_{\rm int} \label{1}\ee
 where $H_S$ and $H_A$ denote the free Hamiltonians of the system and
apparatus respectively, and $H_{\rm int}$ is the interaction Hamiltonian.  
(As usual, $H_S$ and $H_A$ are shorthand for the operators $H_S \otimes
\openone$ and $\openone\otimes H_A$.)
 Further, $g(t)$ is a switch function, which takes a constant value 
$1/\tau$  during a very long interval $[0,\tau]$, and vanishes
smoothly and rapidly before and after this interval.
We take
 \be  H_{\rm int} = O \otimes P \ee
where $O$ is an observable of the system and $P$ is the canonical momentum 
conjugate to the pointer position of the apparatus. 

In general, the evolution generated by a time-dependent  Hamiltonian
such as (1) is given by
\be U =
{ \cal T} e^{-i   \int H_{\rm tot} (t)\,dt} \ee
 where ${\mathcal T}$ is the time-ordering operator.  In the present case,
the Hamiltonian  commutes with itself  during the period $[0,\tau]$ 
(i.e.\ $ [H_{\rm
tot}(t) , H_{\rm tot}(t')] =0$ for $0\leq t, t' \leq \tau$), so that
time-ordering is unimportant
during this period.
If we neglect  the small interaction during the switching  on and
off
periods
we can  write:
  \be
U=   e^{ -i \bigl( \tau(H_S + H_A) + O\otimes P\bigr)} 
.
 \label{5}\ee  

We further need two specific assumptions about the Hamiltonian (1). First,
it is assumed that the system Hamiltonian is completely non-degenerate
and discrete.
Thus, $H_S$ has a complete orthonormal set of non-degenerate eigenvectors
 $\{ \ket{\phi_1}, \ket{\phi_2}, \ldots \}$ in ${\mathcal H_S}$ such that
   \be H_S\ket{\phi_n} = E_n \ket{\phi_n}.\ee
 Secondly, we will assume that the pointer momentum $P$ commutes with the
free Hamiltonian of the apparatus, i.e.:
   \be [H_A , P ] =0\ee
Let now the 
initial state be of the form:
\be \ket{\Psi}_{\rm i} = \ket{\phi_n}\ket{\chi} \ee
where $\ket{\chi}$ is an arbitrary normalized state in ${\cal H}_A$.
The evolution (\ref{5}) will transform this into \be
\ket{\Psi}_{\rm f}
= U\ket{\Psi}_{\rm i} = 
 e^{-i \bigl( \tau(H_S  +H_A) + O\otimes P\bigr)} \ket{\phi_n} \ket{\chi}.
\label{9}\ee 
 At this point we note that since $g$ is slowly varying and $|g(t)|\ll 1$
during the entire interval $[0,\tau]$, one can apply the adiabatic theorem
and first order perturbation theory. I will not go into the details of
these approximation theorems but merely note the result:
for the special choice (8),  and in the
limit
$\tau\rightarrow
\infty$,
the following  approximation for  the final state obtains:
\be \ket{\Psi}_{\rm f}  \approx e^{-i \tau  E_n } \ket{\phi_n}\, e^{-i(\tau
H_A
 +
\midd{O}_n P) } \ket{\chi} \label{10}\ee
where 
\be \midd{O}_n = \bra{\phi_n} O \ket{\phi_n}.  \ee
(See refs. \cite{AAV,Dass} for more details.)

The result (\ref{10})  has two important features: First, 
 the apparatus state has been changed, not only by the free evolution
$e^{-i \tau H_A}$,  but also by the additional action of the operator
 $e^{-i\midd{O}_n P}$. 
Since  $[H_A, P]=0$, this second operator  
 corresponds to a shift, proportional to $\midd{O}_n$, in the pointer
position
variable $Q$ that is canonically conjugate to $P$. 
 Thus, if $e^{-i\tau H_A}\ket{\chi}$ is
a state such that the pointer position is reasonably well-defined 
 (i.e.\ the wave packet $\langle q |e^{-i\tau H_A} \ket{\chi}$ vanishes
outside of an interval smaller than $\min_{n,m} |\midd{O}_n -
\midd{O}_m|$) we can infer the value of $\midd{O}_n$ with certainty by a
reading of $Q$ after the interaction is over. Thus, the measurement
procedure indeed yields the expectation  value
of the observable  $O$   of the system.

The second important feature is that there is no entanglement in the final
state (\ref{10}). That is to say, we can read the value of
$\midd{O}_n$ from the pointer position
without causing a collapse or reduction of the system-plus-apparatus
state vector.  In fact, the state vector of the system 
changes merely by a phase factor. Since state vectors differing by
a phase  factor represent the same state, this means that the state
of the system remains completely unchanged in this procedure.
Hence the name `protective measurement'.

This is in sharp contrast to the usual von Neumann measurement scheme
where measurement of an arbitrary observable generically leads to
entanglement. In this case the subsequent reading (or permanent
registration) of the pointer observable leads to disruption of the
coherence of this entanglement, because of the projection postulate. In a
protective procedure, however, the system is still available in its
original state after the measurement.
 We can then repeat the procedure for other observables $O', O'',
\ldots$
and determine their expectation values as well. 
 Continuing in this manner, one eventually obtains sufficient data to
determine the exact quantum state of the system uniquely (up to an overall
phase factor). E.g.\ if $\mathcal H_S$ is two-dimensional, three linearly
independent observables will suffice.
 This is the basis for the conclusion of \AAV{} that one can observe the
state by means of protective measurements on an individual system.

Note that this conclusion is obtained only under the condition that the
system
was initially in an eigenstate of $H_S$. Indeed,
as pointed out by \AAV,  if the system is initially described by  a
superposition $\ket{\psi} =\sum_n c_n \ket{\phi_n}$ 
 the protective measurement brings about the
evolution \be
  \sum_n c_n \ket{\phi_n}\ket{\chi} \longrightarrow 
  \sum_n c_n e^{-i\tau E_n} \ket{\phi_n}
e^{-i(\tau H_A +
\midd{\tilde{O}}_n P)} 
\ket{\chi}  \label{29}\ee
which results  in an entangled superposition,
just as in the von Neumann
measurement.

However, they argue that in this case a protective measurement may still
be feasible, by tailoring the Hamiltonian (e.g.\ by applying external
fields) such that $\ket{\psi}$ becomes a non-degenerate eigenstate.

\section{Consequences for the meaning of the quantum state}

We have seen that by means of a protective measurement, the
expectation values $\midd{O}_n$ can be obtained, for arbitrary $O$, for an
individual system in an eigenstate  $\ket{\phi_n}$  of the Hamiltonian
$H_S$, without altering this state.
By a sequence of protective measurements it is then 
possible to completely determine the quantum state of 
that system. 
 What does this entail for the interpretation of the quantum state?
Aharonov and Vaidman write:
     \begin{quote} ``We have shown that stationary quantum states can be
observed.  This is our main argument for associating physical reality with
the quantum state of a single particle.''\cite{AV}
      \end{quote}
Indeed, there is immediate intuitive appeal for this conclusion.
If the value $\midd{O}_n$ can be determined by inspection of a single
particle, it seems natural to assume that the particle somehow `knows' 
this value, i.e.\ that it corresponds to a real attribute.  
When this holds for arbitrary  observables $O$,  all their
expectation values
represent real attributes. But this is equivalent to assuming that the
quantum state  itself is a real property of the particle.

Note, however, that this claim 
 is established only for the eigenstates of $H_S$.  That is to say, 
the protective measurement determines the state of a single system
without disturbance  only if it is
given beforehand that this state belongs to the orthonormal family $\{
\ket{\phi_1}, \ket{\phi_2}, \ldots\}$.
 One might object that under this restriction
 the achievement is not surprising.  Indeed, the same claim could be made
for a traditional Von Neumann measurement of $H_S$. In that case, the
measurement brings about the evolution
 \be
   \ket{\phi_n}\ket{\chi_0} \longrightarrow 
 \ket{\phi_n}
\ket{\chi_n} \label{vn}
 \ee for a special initial state $\ket{\chi_0}$ and with $\ket{\chi_n}$
denoting orthonormal pointer states. One can then identify $\ket{\phi_n}$
simply by reading off the label $n$ from the pointer state.

Anticipating this type of objection, \AAV{} emphasize \cite{AAV,AAV2} that
 in a protective measurement, one does not need knowledge of $H_S$.
 Thus, while in the above Von Neumann measurement one can only reconstruct
$\ket{\phi_n}$ from the observed value of $n$ if the Hamiltonian (or the
set of its eigenstates) is known, the protective measurement yields the
data $\midd{O}_n, \midd{O'}_n$ etc.  From this, one can reconstruct the
form of $\ket{\phi_n}$ even when $H_S$ or the precise form of its
eigenvectors are unknown.
 Thus, even if the information obtained in a protective measurement is of
the same kind as that in a Von Neumann measurement,
it is obtained under different conditions.

One may still doubt whether this rebuttal is convincing.  Obviously, in a
protective measurement, the reconstruction of the state from the
experimental data requires knowledge of the exact form of
the observables $O,O'$ etc. This requirement seems completely analogous to
the condition that the $H_S$ is known in the case of the Von Neumann
measurement.

However, even if one concludes that the claim that it is possible to
determine state of an individual system, under the condition that this
state is a member of some orthonormal set is not by itself spectacular,
 the fact remains that a protective measurement achieves this result in a
surprising manner. The procedure records the expectation value of an
observable $O$ in a single measurement, while the system is not
necessarily in an eigenstate of $O$. This could never be achieved in a Von
Neumann measurement

Another type of objection concerns the inference in the above argument
 from observability to physical reality of the state. Dickson
\cite{Dickson} points out that this argument will not carry appeal for
those who adopt an instrumentalist view, i.e.\ for those who regard the
theory as merely a recipe for predicting experimental results; or for
empiricists who may accept a theory when it is empirically adequate
without feeling committed to believe any of its ontological claims.

However, one can also use the argument, not so much to infer the reality
of the quantum state, but rather as a weapon to attack the internal
coherence of other interpretations. Indeed, suppose some interpretation of
quantum mechanics
   denies or qualifies the unconditional existence of the wave function of
an individual system-- as in fact most interpretations do. It would then
seem most surprising, to say the least, if one can still determine its
exact form by measurements on such a individual system.
 Hence, even if one doubts whether the analysis establishes the physical
reality of the wave function, it can still be effective in establishing
incoherence of alternative interpretations. 

   Let me mention two of them. The most obvious candidate in danger from
the above conclusions is the `ensemble' or statistical interpretation. In
this view, adopted by authors such as Einstein, Popper, Blochintsev and
Ballentine, the quantum state describes not an individual system, but
rather an ensemble of systems.
 The quantum mechanical expectation values are then interpreted as
averages over the members of the ensemble. Accordingly, one cannot
determine the state of an individual system, simply because it is not a
property of a single system.  This view appears untenable in the light of
the above claims.

Secondly, in the Copenhagen view, the quantum state is assumed to give a
complete description of the individual system. But this description is,
according to Bohr, `symbolic' and does not literally represent physical
reality in the sense of a one-to-one correspondence.  While the quantum
state encodes complete information about the system, only part of this
information is applicable in any given measurement context.  Due to the
principle of complementarity, one always has to collect experimental data
from mutually exclusive measurement arrangements to obtain a full
determination of the state.

To be more precise, a measurement context in which the non-degenerate
observable $A$ is measured is represented by  the eigenbasis of the 
observable. In this basis we can expand the state, say
 $\ket{\psi} = \sum_i c_i \ket{a_i}$
 and the $|c_i|^2$ give the probabilities of finding the outcomes
$a_i$.
 The phase relations between the coefficients $c_i$ are not accessible in
this context. To determine them, one needs to consider a measurement of
some other observable $B$ that does not commute with $A$.
 But the context defined by $B$ is, according to the
Copenhagen point of view, incompatible with the original one.
 Therefore, we can never obtain sufficient information to determine the
quantum state of an individual system in a single context.
By contrast, the series of protective measurements needed for a
determination of the state are not mutually exclusive.  Thus the claim of
\AAV{} amounts to nothing less than a disproof of the principle of
complementarity.

Remarkably, some other interpretations are no better off, even if they
agree with
the point of view that the wave function of an individual  system 
represents  a physically real entity. 
 For example, in the Bohm
interpretation, the modulus  $R(x)$ of  the wave function 
 $\psi(x)= R(x)e^{iS(x)/\hbar}$
 appears in the quantum potential $U(x) = \frac{- \hbar^2}{2m}
\frac{\Delta R}{R}$ which represents an independently existing potential
acting on the particle. In some versions of the Bohm interpretation, the
phase $S(x)$ represents a real entity as well (the `guidance field').  
Nevertheless, it has been shown that these fields cannot be determined
experimentally from the behavior of an individual particle
\cite[p.~369-378]{Holland}.  Hence, the Bohm interpretation is also
committed to the conclusion that one cannot observe the wave function by
inspection of an individual particle.

It follows that in order to avoid damage for the above interpretations,
one should question the very starting point of the above argument, i.e.\
whether it has been sufficiently established that it is possible to
perform a protective measurement for arbitrary observables. 
 In the next section we shall see that there are indeed severe 
restrictions
on the performance of protective measurements. 

\section{Restrictions on the observables that can be measured
protectively}

It is essential to note that for the purpose of the protective measurement
the form of the evolution obtained in (\ref{10}) should hold for all
$\ket{\phi_n}$. Indeed, we are assuming that all we know about the initial
state of the system is that it is one of the eigenstates of a
non-degenerate Hamiltonian $H_S$, but not which one.  It is the purpose of
the procedure to determine this state.
 Therefore, one must guarantee that the desired form of the evolution
holds for all $\ket{\phi_n}$, i.e.\ the approximation must be a good one
for all these states. We exploit this to derive a simple but very
restrictive property of the evolution.

Let us  define an operator  $U_{\rm app}$ that brings about 
the approximate  evolution 
(\ref{10}) exactly for all vectors of the form
$\ket{\phi_{n}}\ket{\chi}$,
i.e.:
\be U_{\rm app} \: :\:
 \ket{\phi_n} \ket{\chi} 
 \longrightarrow  
 e^{-i\tau
E_n}
\ket{\phi_n }
 e^{-i (H_A \tau +  \midd{O}_n P)}  
 \ket{\chi} .  \label{12}
\ee
 By linearity this extends to a unique definition of $U_{\rm app}$ as an
operator on ${\cal H}_S\otimes{\cal
H}_A$.
 One can also  give an explicit expression for 
 $U_{\rm app}$.
Let 
\be\tilde{O} = \sum_n P_n OP_n \label{tilde}\ee
be an operator  on $\mathcal H_S$, where $P_n = \proj{\phi_n}$.  It
is easy to see
that
 \be U_{\rm app} = e^{-i (H_S + H_A)\tau -i \tilde{O}\otimes
P}\label{30} \ee
 by checking that the  right-hand side  indeed produces the transition
(\ref{12}) when acting on states of the form
$\ket{\phi_n}\ket{\chi}$.

   But then, since $[\tilde{O}, H_S]=0$, it immediately follows that
$[U_{\rm app},H_S]=0$, or in other words:
\be
 U_{\rm app}^\dagger
H_S  U_{\rm app}  
= H_S 
 \label{15} .
\ee
 This means that $H_S$ is conserved under the evolution $U_{\rm
app}$.

 This already suggests that the observable  $O$ appearing in
the interaction Hamiltonian of a
protective measurement must be subject to restrictions.
 Indeed, if the evolution operator $U$ given by equation (\ref{9})
contains an
arbitrary self-adjoint operator
  $O$, one would not expect that,  to good approximation,
$U$ commutes with $H_S$. One would suspect that this is the case only
if $O$ commutes with $H_S$. 
 However, since $U_{\rm app}$ is only an approximation to $U$, we have to
be careful to spell this suspicion  out.

To say that the  approximation involved in equation (\ref{10}) is good
means that 
\be  \label{17}
 \| (U - U_{\rm app} ) \ket{\phi_n}\ket{\chi} \|
\rightarrow 0 \; \mbox{ if }\; \tau \rightarrow \infty .\ee
As mentioned earlier,  we assume  this  holds for 
for all $n$.
  Moreover, the 
approximation theorems apply for arbitrary $\ket{\chi}$.
Thus, (\ref{17}) holds  
for all $\ket{\phi_n}$ and 
$\ket{\chi}$.
This condition is then equivalent to 
\be
 \lim_{\tau \rightarrow\infty} \| U - U_{\rm app}   \| =
0.\ee
Together with (\ref{15}) this implies\footnote{%
            Because   
	$\|U^\dagger H_SU -H_S\| =\|U^\dagger H_SU -
	U^\dagger_{\rm app} H_SU_{\rm app}\|
                         = \|(U- U_{\rm app})^\dagger H_SU +
                            U^\dagger_{\rm app} H_S(U- U_{\rm app})\|
               \leq 2
              \|U- U_{\rm app}\|  \, \|H_S\|\rightarrow 0$, 
          at least if $H_S$ is bounded.  However, if $H_S$ is unbounded, 
          the argument can be rerun, while replacing $H_S$ with
             the set of its spectral
                      projections.}
 \be \lim_{\tau \rightarrow \infty} \| U^{\dagger} H_SU -H_S \|= 0
.\label{20}\ee

Now consider the matrix element
\be 
\bra{\phi_m}
\bra{\chi}
\left( U^{\dagger} H_SU  -H_S\right) \ket{\phi_n}\ket{\chi}
\label{21}.\ee
 Since, for any self-adjoint operator, the operator norm majorizes the
absolute value of its matrix elements, we conclude from (\ref{20}) that, 
as
$\tau \rightarrow \infty$,
 \be 
\bra{\chi}
\bra{\phi_m}
 U^{\dagger} H_SU  \ket{\phi_n}\ket{\chi} \longrightarrow 
\bra{\chi}
\bra{\phi_m}
H_S\ket{\phi_n}\ket{\chi} =  E_n \delta_{nm}. \label{22}
 \ee
	Let $\{\ket{p,\alpha}\}$ be a complete orthonormal set of
(improper) common eigenstates in $\mathcal H_A$ of both $H_A$ and $P$:
  \be P \ket{p} = p \ket{p,\alpha}, \;\;\; H_A \ket{p} = E(p,\alpha)
\ket{p, \alpha}.
\ee
Here, the index $\alpha$ is used to allow for degeneracy in $P$ and $H_A$.
We  expand  the left-hand side of (\ref{22}):
\BE \lefteqn{
\bra{\chi}
\bra{\phi_m}
 U^{\dagger}H_S U\ket{\phi_n}\ket{\chi}
=}\\&&
\sum_{\alpha \beta} \int\!\!\int dp dp' \,
\inpr{\chi}{p',\beta}
\bra{p',\beta} 
\bra{\phi_m}
U^\dagger H_S   U
\ket{\phi_n}\ket{p,\alpha}
\inpr{p,\alpha}{\chi}
\nn\\
&=&
\sum_{\alpha \beta} \int\!\!\int dp dp'\,
\inpr{\chi}{p',\beta}\inpr{p,\alpha}{\chi} \times \nn
   \\&&
\times \bra{p',\beta} 
\bra{\phi_m}
e^{i \bigl(\tau (H_S + E(p',\beta)) + p'O \bigr)} H_S 
 e^{-i \bigl(\tau (H_S + E(p,\alpha)) +p O \bigr)} 
\ket{\phi_n}\ket{p,\alpha}\nn\\
&=& 
\sum_{\alpha \beta} \int\!\!\int dp dp'\,
\inpr{\chi}{p',\beta}
\inpr{p,\alpha}{\chi}
\inpr{p',\beta}{p,\alpha} \times \nn\\
&& \times e^{i\tau(E(p',\beta)-E(p,\alpha)} 
 \bra{\phi_m} 
e^{i(\tau H_S +p'O)} H_S e^{-i(\tau H_S +pO)}
\ket{\phi_n} 
\nn\\
&=& 
\sum_{\alpha } \int dp \,
|\inpr{p,\alpha}{\chi}|^2
\, 
 \bra{\phi_m} 
e^{i(\tau H_S +pO)} H_S e^{-i(\tau H_S + pO)}
\ket{\phi_n} 
 \label{24}
\EE
According to (\ref{22}) this expression will approach zero if $m\neq n$
and 
$E_n$ otherwise. 
 But, since $\ket{\chi}$ is arbitrary, this happens only if for almost all
values of $p$:
 \[
 \bra{\phi_m} 
e^{i(\tau H_S +pO)} H_S e^{-i(\tau H_S + pO)}
\ket{\phi_n} 
 \rightarrow E_n \delta_{mn}
 \]
or equivalently\[
e^{ i\tau(E_m- E_n)}
 \bra{\phi_m} 
e^{ipO} H_S e^{-i pO}
\ket{\phi_n}   \rightarrow E_n \delta_{mn}.
  \]
This means that for almost all $p\in I\!\!R$,
\be e^{ipO} H_S e^{-ipO}  =  H_S, \ee
which implies:
\be [O, H_S ] =0\ee
 Thus we  conclude that $U_{\rm app}$ is a good approximation
to $U$ only if the observable $O$ commutes with the system Hamiltonian.

Notice that we did not rely on the differential form of the evolution
(\ref{5}). 
Had we done so, we would have immediately  obtained the result 
  \be [H_S + H_A + g(t) O\otimes P , H_S] = g(t) [ O,H_S] \otimes P
\rightarrow 0   \label{27}\ee
by noting that
 the switch function $g$ 
is of the order 
$g \approx \tau^{-1}$ so that the commutator (\ref{27}) vanishes
automatically in the
limit 
$ \tau\rightarrow \infty$. Thus, 
this approach would not reveal a constraint  on $[O, H_S]$. 

  \section{An alternative look at protective measurements}
 
We have reached the conclusion that the assumptions involved  in 
a protective measurement entail that the observable whose
expectation value is obtained commutes with the Hamiltonian $H_S$ of the
system. This obviously presents a major restriction.
 In Copenhagen terms, it means that the information provided by a
protective measurement is restricted to that belonging to a single
measurement context only. Indeed, in view of this, one might even doubt
whether the claim that the quantum state can be uniquely determined by
means of protective measurement is valid at all. I shall argue here that
this claim is still true, but at the same time, that this need not be
interpreted as evidence for the physical reality of the quantum state.

To see this, let us compare the approximative evolution (\ref{tilde}) with
 the (almost) exact evolution (\ref{5}).
 This shows that the approximations involved 
 amount to the replacement of the original
observable $O$ by the sandwiched observable $\tilde{O}$.  But this
observable combines two interesting virtues:  (i) it commutes with $H_S$
and (ii) its expectation value in any eigenstate $\ket{\phi_n}$ equals
that of $O$: $\midd{ \tilde{O}}_n = \midd{O}_n $. Thus the measurement of
$\tilde{O}$, which is compatible with $H_S$,  suffices to determine the
value of $\midd{O}_n$.

Thus we can give an alternative explanation  for what happens in a
protective measurement, which does not appeal to the idea that
an individual system carries  information about its quantum
state.
 The interaction between system and
apparatus is produced by a very small
interaction term, viz.\ $g(t)O\otimes P$, that works for a very long time.  
The smallness is responsible for the fact that $\ket{\phi_n}$ remains
unchanged, the long time explains that nevertheless a non-vanishing effect
of the interaction builds up in the state of the apparatus.  However, the
effect that builds up in the course of time is due only to the part of $O$
(namely $\tilde{O}$) that commutes with $H_S$.  It is only this operator
whose expectation value is revealed. The procedure is insensitive,
however, to to the remainder $O - \tilde{O}$, i.e.\ the part of $O$ that
does not commute with $H_S$.
 In fact, this statement can indeed be immediately verified: if we replace
$\ket{\phi_n}$ in the initial state with an arbitrary superposition
of the form $\sum_n c_n \ket{\phi_n}$,
 the protective measurement brings about the transition (\ref{29}).
Here a reading of the pointer variable
 invariably leads to a disruption of the  coherence of the terms,
and we are cut off from establishing the phase relations between the
coefficients $c_n$.
In \AAV's terminology, this is expressed by saying that superpositions
of
the eigenstates $\ket{\phi_n}$ are not protected in this particular 
procedure.
 But from the present point of view,
 the incapability of a protective
measurement to reveal the phase relations in a superposition, i.e.\
the incapability of
discriminating the superposition 
 $ \sum_n c_n \ket{\phi_n}$
 from the corresponding mixture 
$  \sum_n |c_n |^2 \ket{\phi_n}\bra{\phi_n}$,
can also be interpreted by saying that one is actually measuring the
observable $\tilde{O}$
rather than $O$.

In short, an alternative explanation for the surprising features of a
protective measurement is that when  one enforces  the adiabatic
conditions, i.e.\ the validity of 
 the approximation
(\ref{10}), the 
observable $O$ is effectively replaced by $\tilde{O}$.  This has no effect
on its
expectation value in the eigenstates $\ket{\phi_n}$ but a large effect on
its commutation relation with $H_S$.

Let us try to illustrate these conclusions by means of an example.
Perhaps one of the most striking examples discussed in ref.~\cite{AAV} is
that of a charged particle (say a proton), which is  described by
a superposition of two states
localized in
distant boxes $L$ and $R$: 
\be 
\ket{\phi_+} =  \frac{1}{\sqrt{2}}(
\ket{\phi_L} +
\ket{\phi_R} ) \label{+}\ee
where 
$\ket{\phi_L}$ and 
$\ket{\phi_R}$ are the ground states of the box potentials. 
The question is whether one can 
demonstrate  that the proton is in this delocalized state.

If the two boxes are bordered by  infinite potential walls, the state 
(\ref{+}) is degenerate with 
\be 
\ket{\phi_-} =  \frac{1}{\sqrt{2}}(
\ket{\phi_L} -
\ket{\phi_R} ) \label{-}\ee
so that the analysis of section 2 would not be applicable. 
But if one arranges  that  in the region between the two boxes   the
potential  has a   large but finite constant value $V$, the states 
$\ket{\phi_L}$ and 
$\ket{\phi_R}$  develop small tails into this middle region, and 
one  achieves   that $\ket{\phi_+}$ and 
$\ket{\phi_-}$ are no longer degenerate. (See Figure 1.)

Now suppose we  measure the position of the proton, 
or somewhat more crudely,  the observable:
\be O = 
 - \proj{\phi_L} + 
\proj{\phi_R} \label{Proj}\ee
This can be done be sending a charged test particle, e.g. an electron,
straight through the
middle between the boxes, perpendicular to the line
joining the two boxes and observing whether its trajectory
 deviates  from a
straight line. 
 \AAV{} show that if the procedure is that of a
conventional
 Von Neumann measurement, one will find a deflection of the electron to
the left or right, with equal probability. Therefore, this 
 procedure does not yield evidence that the proton is in 
the delocalized  state $\ket{\phi_+}$.

However if the measurement is protective, the result is very different. 
The trajectory  of the electron  is now only sensitive to $\midd{O}_+ =0$
and, therefore, it will continue through the boxes  without deviation. 
This then  seems a clear demonstration that the 
proton is really in
a delocalized 
superposition. In the words of  ref.~\cite{AAV}, 
``the interaction is as if
half of [the particle] is in box [L] and the
other is in box [R].''  and:
``the protective measurement shows the manifestation of the wave function
as an extended object.''

How should one analyze this example from the point of view proposed 
above? In this view, 
the protective measurement does not  measure $O$, 
but rather  a related
observable 
$\tilde{O}$.
If, for simplicity,  we restrict ourselves  to the two-dimensional Hilbert
space spanned by  
$\ket{\phi_+}$  and 
$\ket{\phi_-}$, an easy calculation shows that in this example:
\be \tilde{O}  = 
 \sum_{j\in \{+,-\} } \proj{\phi_j} O
\proj{\phi_j}
=0 \ee
 This means the null result of the experiment should not surprise us:  
this particular protective measurement is incapable of yielding any other
result!

This conclusion  can be straightforwardly verified  by considering 
 the  case where the procedure is carried out on a proton 
 prepared  in a localized state, say  $\ket{\phi_L}$.
Since this state is not protected in the procedure,
 one  obtains the evolution:
\be 
\ket{\phi_L}\ket{\chi}=
\frac{1}{\sqrt{2}} \left( 
\ket{\phi_+} + 
\ket{\phi_-} \right) 
\ket{\chi}
\rightarrow 
\frac{1}{\sqrt{2}} \left( 
\ket{\phi_+}\ket{\chi_0} 
+
\ket{\phi_-}
\ket{\chi'_0} 
\right)
\label{loc} \ee
where 
$\ket{\chi'_0}$
 is the  final state of  the electron  in the case when 
the proton was initially in the state
$\ket{\phi_-}$. Since 
$
\midd{O}_+ =
 \midd{O}_- = 0$, the electron  travels a straight trajectory in
 the
state $\ket{\chi'_0}$ as
 well  as in 
$\ket{\chi_0}$.\footnote{There may be slight distinction between
	$\ket{\chi_0}$ and $\ket{\chi'_0}$  because of a different
	acceleration experienced by the electron, due to the different
	shape of the tails of $\ket{\phi_{\pm}}$ in the region between the
        boxes.
	However, in the adiabatic limit,
	this distinction will disappear.  
    }
 Thus,  the electron  will indeed travel on a   straight
path, 
 regardless of whether the proton is delocalized or not.
  Therefore,
this  experiment provides no evidence for the spatial
delocalization of the  proton.

On first sight, the conclusion that the electron is not deflected, even if
the proton is localized, may seem counterintuitive because of the
asymmetry of the Coulomb field produced in this case.
 But note that the adiabatic limit in this experiment involves letting the
distance between the boxes, and the value of the potential in the middle
region go to infinity. Consequently, the electrostatic force on the
electron, and hence the curvature of its trajectory trajectory also
vanishes in this limit.

\section{Conclusion and discussion}

It has been shown here that for a system with a non-degenerate free
Hamiltonian $H_S$, a protective measurement is only possible of
observables $O$ that commute with $H_S$.
 This is not in conflict with the claim that the measurement procedure
is able to 
yield the expectation value of an arbitrary observable $O$. The
explanation is simply that in the regime in which the conditions and
approximations for the adiabatic theorem and first order perturbation
theory are valid, the procedure actually measures another observable
$\tilde{O}$ which commutes with $H_S$, but
for the considered set of states, has the same expectation value
as $O$.
 A similar conclusion has been reached by Rovelli~\cite{Rov} by
analysis of a concrete example. 

In this explanation we do not need recourse to a manifestation of the wave
function in the individual system. Rather, it is clear that in a
protective measurement we are dealing with what from the Copenhagen point
of view would be characterized as a single measurement context only: that
of the Hamiltonian $H_S$. All the information obtained is in fact
compatible with this context. Hence there is no threat to the
complementarity principle.

Similarly, the ensemble interpretation of the wave function can be saved
from incoherence. Assume that $\ket{\phi_n}$ describes an ensemble of
similarly prepared systems. The ensemble is dispersionless for the
Hamiltonian, and hence all members will produce identical outcomes when
$H_S$ is measured. The same holds for a measurement of $\tilde{O}$:
since $\tilde{O}$ is a function of $H_S$, and $\ket{\phi_n}$ is its
eigenstate, with
  \be \tilde{O} \ket{\phi_n} = \midd{O}_n \ket{\phi_n} \ee all the members
of the ensemble will therefore reveal the same value $\midd{O}_n$ in the
measurement of $\tilde{O}$. It is not necessary to conclude that,
paradoxically,
 an
individual  
 system carries complete information about the quantum state, i.e.\
that it `knows' to
which ensemble it belongs.

Finally, I want to discuss two  possible objections to the present
conclusions.
 First,  an essential assumption I have used in section 4 is that  the
approximation 
(\ref{10}) is valid for all states  $\ket{\phi_n}$.
However, one may object this is too
restrictive.  A protective
measurement might still be  of interest, also  if 
the approximation is valid only for some subset, call it $J$, of $\{ 
\ket{\phi_1}
,\ket{\phi_2}, \ldots \}$.

In that case, the procedure would allow us to determine the state only when it
is given that the  initial state belongs to the subset $J$.
However,  since $H_S$ is
conserved  in  the subspace  spanned by the set $J$,
the final state  lies in the same subspace.
  Thus,
effectively, it then suffices to restrict our attention to a reduced
Hilbert
space, spanned by $J$.
 But in this reduced space we can give the same argument as above, because the
approximation will now by valid for all eigenvectors of $H_S$ in the reduced
space.  Hence, this escape route will not not bring any essential change in our
conclusions.

A second objection may be that I have not discussed the possibility of
changing $H_S$ between two protective measurements (e.g.\ by applying or
varying some external fields). Indeed, one can imagine that a first
protective measurement measures an observable $O$ which commutes with
$H_S$ and then the Hamiltonian is changed to $H'_S$ whereafter an
observable $O'$ such that $[O', H'_S]=0$ is measured protectively, etc.

Thus, if $[H, H'] \neq 0$, we might still be able to combine information
from incompatible measurement contexts into one experiment.  The problem
with this proposal is of course that one must take care of what happens to
the state of the system.  If $H_S$ is changed abruptly, the system will
generally not be in an eigenstate of $H'_S$ at the start of the second
measurement.  On the other hand, if  $H_S$ is changed quasi-statically, so
that the adiabatic theorem is applicable, one can arrange that the
system's state will transform into an eigenstate of the new Hamiltonian.

A more careful analysis than offered here is necessary to decide whether
such a proposal would lead to a refutation of the complementarity
principle or whether one can still maintain that this measurement defines
a single but time-dependent context. In any case, this proposal would
differ from that of \AAV{} in the sense that here not the protectiveness
\emph{of} the measurements but also  \emph{in between} measurements is
essential.

\section*{Acknowledgements}
 I thank Jeeva Anandan, Jan Hilgevoord, Dennis Dieks, Janneke van Lith and
Erwyn van der Meer and also the inquisitive audience at the Subfaculty for
Philosophy of Oxford University for inspiring discussions and helpful
comments.

\pagebreak
\pagestyle{empty}
\begin{figure}[h]
  \begin{center}
\begin{picture}(288,233)(10,20)
\put(100,74){\vector(0,1){34}}
\put(100,54){\vector(0,-1){33.8}}
\put(100,64){\makebox(0,0){$V$}}
\put(40,48){\makebox(0,0){$\ket{\phi_+}$}}
\put(30,10){\makebox(0,0){$L$}}
\put(258,10){\makebox(0,0){$R$}}
\put(40,100){\makebox(0,0){$\ket{\phi_-}$}}
 \epsfig{file=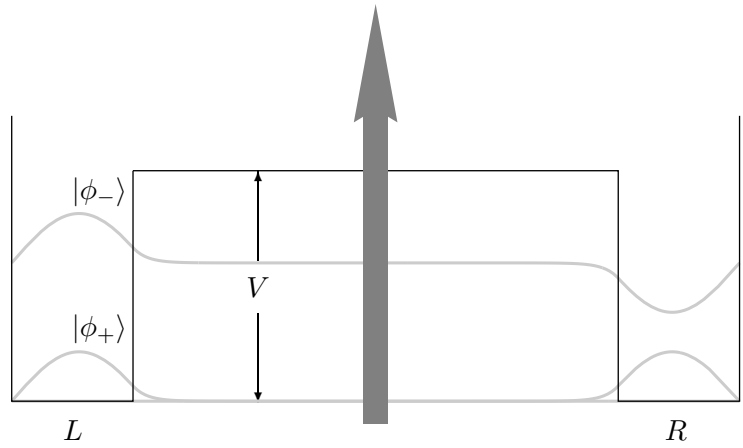}
\end{picture}
  \end{center} \caption{A proton is in superposition $\ket_{\phi_+}$ of
two states localized in boxes $L$ and $R$. In between the boxes there is
an external constant potential $V$ which lifts the degeneracy of
$\ket{\phi_+}$ and $\ket{\phi_-}$. When the location of the proton is
measured protectively, by sending an electron through the middle between
the boxes, the electron will pass the boxes on a straigt trajectory. }
\end{figure} 

\end{document}